\documentclass[journal=jctcce, layout=twocolumn, manuscript=article]{achemso}

\usepackage[T1]{fontenc}
\usepackage{hyperref}
\usepackage{amsmath}
\usepackage{amssymb}

\newcommand*{\ket}[1]{\ensuremath{|#1\rangle}}
\newcommand*{\bra}[1]{\ensuremath{\langle#1|}}

\newcommand*{\fdof}[1]{\ensuremath{\pi_{#1}I_{#1}N_{#1}j_{#1}m_{#1}\alpha_{#1}}}
\newcommand*{\frdof}[1]{\ensuremath{\pi_{#1}I_{#1}N_{#1}j_{#1}\alpha_{#1}}}
\newcommand*{\dof}[1]{\ensuremath{\mathbf{x}_{#1}}}
\newcommand*{\rdof}[1]{\ensuremath{x_{#1}}}
\newcommand*{\dofa}[1]{\ensuremath{\dof{#1}\alpha_{#1}}}
\newcommand*{\rdofa}[1]{\ensuremath{\rdof{#1}\alpha_{#1}}}

\newcommand*{\fdz}[3]{\ensuremath{\delta_{\pi_{#1} + \pi_{#2} \bmod 2, \pi_{#3}}}}
\newcommand*{\dz}{\ensuremath{\delta_\pi}}
\newcommand*{\fdI}[3]{\ensuremath{\delta_{I_{#1} \otimes I_{#2}, I_{#3}}}}
\newcommand*{\dI}{\ensuremath{\delta_I}}
\newcommand*{\fdN}[3]{\ensuremath{\delta_{N_{#1} + N_{#2}, N_{#3}}}}
\newcommand*{\dN}{\ensuremath{\delta_N}}

\newcommand*{\msign}[1]{\ensuremath{(-1)^{j_{#1} - m_{#1}}}}

\newcommand*{\tj}[3]{\ensuremath{\begin{pmatrix}
      j_{#1} & j_{#2} & j_{#3} \\
      m_{#1} & m_{#2} & -m_{#3}
\end{pmatrix}}}

\newcommand*{\order}[1]{\ensuremath{\mathcal{O}\left(#1\right)}}

\newcommand*{\an}[1]{\ensuremath{c_{#1}}}
\newcommand*{\cre}[1]{\ensuremath{c^\dagger_{#1}}}

\graphicspath{{figures/}}

\newcommand*{\cuo}{\ensuremath{[\mathrm{Cu}_2\mathrm{O}_2]^{2+}}}

\title{The Three-Legged Tree Tensor Networks with $SU(2)$- and molecular point
group symmetry}

\author{Klaas Gunst}
\affiliation[Ghent University]
{Center for Molecular Modeling, Ghent University, Technologiepark 46, 
9052 Zwijnaarde, Belgium}
\alsoaffiliation[Ghent University]
{Department of Physics and Astronomy, Ghent University, Krijgslaan 281, S9, 
B-9000 Ghent, Belgium}
\email{Klaas.Gunst@UGent.be}

\author{Frank Verstraete}
\affiliation[Ghent University]
{Department of Physics and Astronomy, Ghent University, Krijgslaan 281, S9, 
B-9000 Ghent, Belgium}
\alsoaffiliation[University of Vienna]
{Vienna Center for Quantum Technology, University of Vienna, Boltzmanngasse 5, 
1090 Vienna, Austria}

\author{Dimitri Van Neck}
\affiliation[Ghent University]
{Center for Molecular Modeling, Ghent University, Technologiepark 46, 
9052 Zwijnaarde, Belgium}

\begin{document}
\begin{tocentry}
  \includegraphics{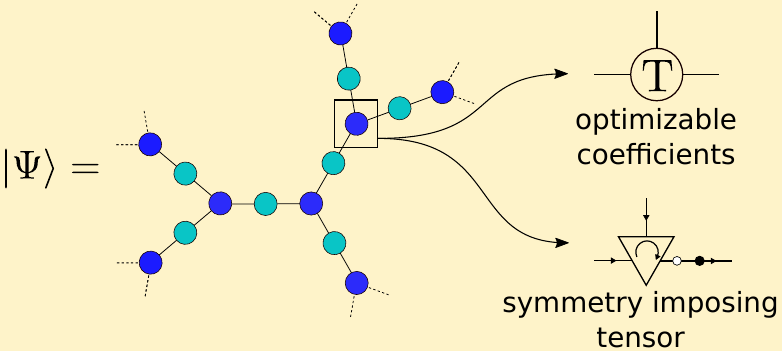}
\end{tocentry}

\begin{abstract}
  We extend the three-legged tree tensor network state (T3NS) [J.  Chem. Theory
  Comput. 2018, 14, 2026-2033] by including spin and the real abelian point
  group symmetries.  T3NS intersperses physical tensors with branching tensors.
  Physical tensors have one physical index and at most two virtual indices.
  Branching tensors have up to three virtual indices and no physical index. In
  this way, T3NS combines the low computational cost of matrix product states
  and their simplicity for implementing symmetries, with the better entanglement
  representation of tree tensor networks. By including spin and point group
  symmetries, more accurate calculations can be obtained with lower
  computational effort. We illustrate this by presenting calculations on the
  bis($\mu$-oxo) and $\mu-\eta^2:\eta^2$ peroxo isomers of \cuo. The used
  implementation is available on github.
\end{abstract}

\section{Introduction}
The discovery of the density matrix renormalization group (DMRG) by S. White in
1992\cite{10.1103PhysRevLett.69.2863, 10.1103PhysRevB.48.10345} introduced a new
and very successful way of treating strongly correlated quantum systems in both
condensed matter physics and theoretical chemistry\cite{10.10631.478295}. Later
on, it was discovered that the DMRG wave function corresponds to a variational
ansatz over the set of matrix product states (MPS)\cite{10.1103PhysRevB.55.2164,
10.1103PhysRevLett.75.3537}. An MPS is a state that can be represented by a
linear chain of tensors. It is the most simple form of a tensor network. The
linear form of the MPS explained the high efficiency of DMRG for systems
respecting the area law for entanglement in one-dimensional quantum spin
systems.

In condensed matter physics, this kick-started the formulation of other
(non-linear) tensor networks suitable for systems with another entanglement
structure than the one-dimensional area law. The projected entangled pair states
(PEPS)\cite{cond-mat0407066v1} and the multiscale entanglement renormalization
ansatz (MERA)\cite{10.1103PhysRevLett.99.220405} are some notable examples.

The linear nature of the MPS is also far from ideal for the entanglement
structure of most molecules.  Hence, other tensor networks have also been
studied in quantum chemistry, like the complete-graph tensor network states
(CGTNS)\cite{10.10881367-26301210103008}, the self-adaptive tensor network
states (SATNS)\cite{10.10631.5004693} and general tree tensor network states
(TTNS)\cite{10.1021ct501187j, 10.1103PhysRevB.82.205105, 10.10631.4798639}.
However, due to its favorable computational complexity and the ease for
exploitation of symmetries, the MPS is still the tensor network method of choice
for quantum chemistry.

Recently, we introduced the Three-Legged Tree Tensor Network
(T3NS)\cite{10.1021acs.jctc.8b00098}, a subclass of the TTNS which allows an
efficient optimization of the wave function while still enjoying the better
entanglement representation of the TTNS. The tree-shaped network allows a
logarithmic growing maximal distance with system size as apposed to the linear
maximal distance for MPS.\cite{10.1021acs.jctc.8b00098, 10.1021ct501187j,
  10.1103PhysRevB.82.205105, 10.1140epjde2014-50500-1, PhysRevB.87.125139,
10.1103PhysRevA.74.022320} In this paper, we will further extend the study of
the T3NS. The implementation of symmetries will be explained and its advantages
will be discussed theoretically as well as illustrated by exemplary
calculations. More particularly, we will use the real abelian point group
symmetries ($C_1$, $C_i$, $C_2$, $C_s$, $D_2$, $C_{2v}$, $C_{2h}$ and $D_{2h}$)
and the $SU(2)$-symmetry present in the non relativistic quantum chemical
Hamiltonian to obtain more accurate and faster calculations. We impose these
global symmetries on the T3NS by using the Wigner-Eckart theorem as introduced
by McCulloch for the MPS\cite{10.10881742-5468200710P10014,
10.1209epli2002-00393-0}.

The paper is structured as follows.  In section~\ref{sec:T3NS}, Three-legged
Tree Tensor Networks and the advantages of this subset of the general tree
tensor network are discussed.  In section~\ref{sec:sym}, the handling of
symmetries in T3NS is illustrated.  We incorporate the parity symmetry
$\mathbb{Z}_2$, the number conservation $U(1)$, the spin symmetry $SU(2)$ and
the real abelian molecular point group symmetries.  We heavily rely on graphical
depictions for the needed tensor contractions, as it yields a concise and clear
notation for otherwise lengthy equations.  In section~\ref{sec:calc}, we show
some calculations of T3NS with the previously mentioned symmetries. We compare
results obtained for \cuo with the ones presented in the original T3NS
paper\cite{10.1021acs.jctc.8b00098}.  Summary and conclusions are provided in
section~\ref{sec:conclusion}.

The used implementation is open source and can be found on Github under the GNU
GPLv3 license\cite{T3NSsourcecode}. 

\section{Three-Legged Tree Tensor Networks}
\label{sec:T3NS}
\begin{figure}[!ht]
  \centering
  \includegraphics[width=\columnwidth]{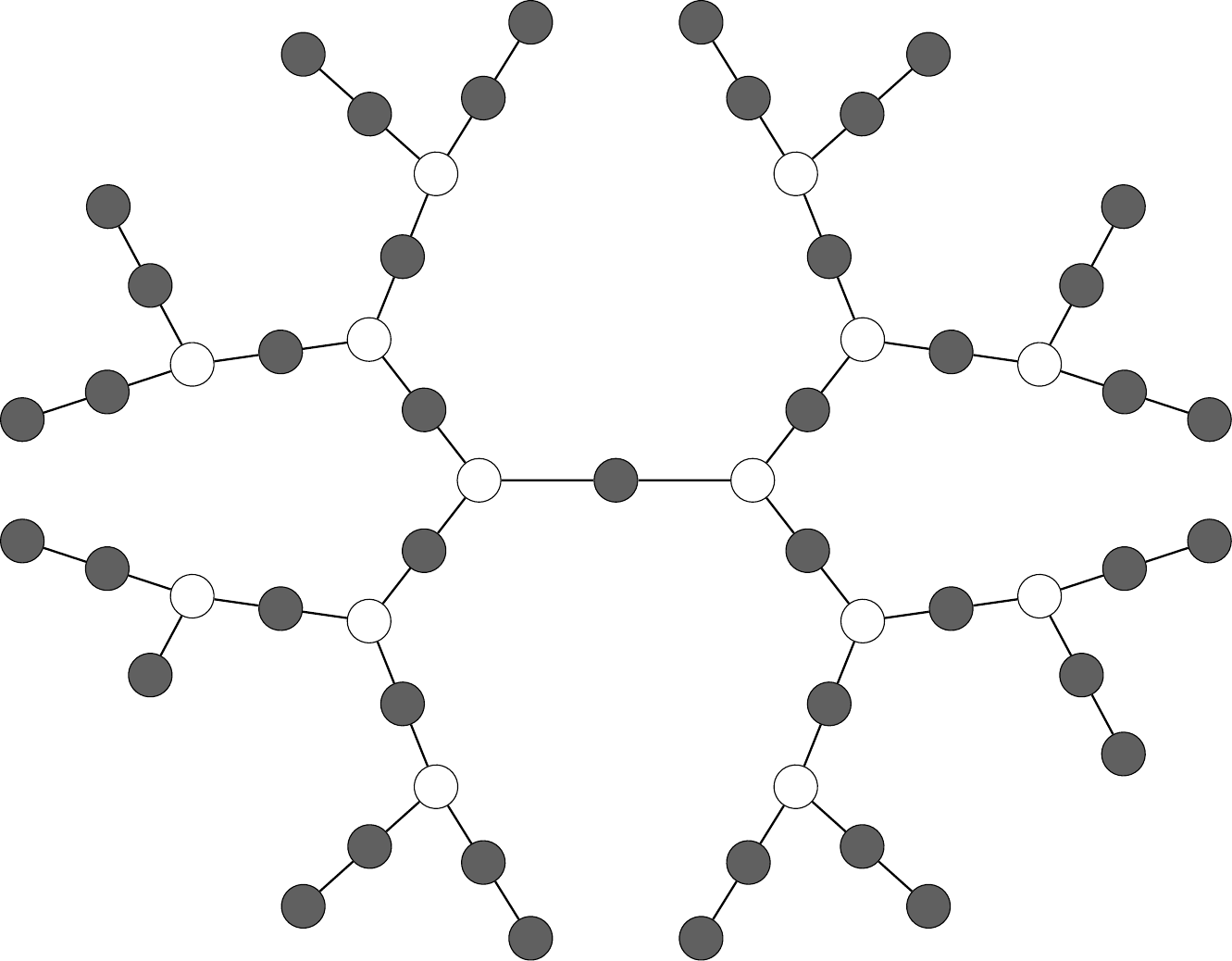}
  \caption{\label{fig:ansatzex}An example of a T3NS ansatz with 44 orbitals.
    Filled circles represent \emph{physical} tensors which have one physical
    bond (not drawn here for simplicity) and maximally two virtual bonds.
    Unfilled circles represent \emph{branching} tensors which have three virtual
    bonds and no physical bonds.
  }
\end{figure}
While the MPS ansatz of DMRG can be represented by a linear chain of tensors,
the TTNS ansatz is built by making a tree-shaped network of tensors, making it
the most general loop-less tensor network state. 

The structure of the TTNS ansatz looks more faithful in the representation of
the entanglement structure of molecules than the linear nature of the MPS
ansatz. However, due to the branching of the network, the complexity for
optimization becomes quickly unfeasible, especially for two-site optimization.
Nakatani and Chan\cite{10.10631.4798639} introduced the half-renormalization for
TTNS to allow efficient two-site optimizations. During the half-renormalization
step the local optimization problem is exactly mapped to an MPS, which is then
optimized.  Although this mapping reduces the complexity per sweep, it also
increases the number of sweeps needed for convergence.

The Three-Legged Tree Tensor Network\cite{10.1021acs.jctc.8b00098} is a subset
of the general TTNS and an other method for circumventing the high complexity of
the TTNS while still maintaining the advantages. In a T3NS, we make use of
\emph{branching} and \emph{physical} tensors for the wave function ansatz and we
intersperse these in the network.  Physical tensors are the same as the tensors
appearing in an MPS: they have one physical bond and maximally two virtual
bonds. Branching tensors have three virtual bonds and no physical bonds. They
allow the tensor network to branch. In a T3NS ansatz, branching tensors are
never placed next to each other, since this would worsen the complexity for
two-site optimization. Similarities exist between the T3NS ansatz and the
half-renormalization of Nakatani and Chan\cite{10.10631.4798639}. The mapping of
the TTNS to an MPS corresponds with a branching tensor. The branching tensor is
never explicitly optimized in the half-renormalization algorithm which results
in the increased number of sweeps needed for convergence. For one-site
optimization, half-renormalization is equivalent with a T3NS sweep where the
branching tensors are never explicitly optimized.  For two-site optimizations,
the translation of half-renormalization to the T3NS is less straight forward but
possible by reshaping the tensor network at each optimization step.  We also
note that a T3NS ansatz with only physical tensors corresponds with an MPS.

Another very important motivation for the restriction to tensors with three legs
is the simplification of the implementation of $SU(2)$-symmetry. This is the
main focus of this paper and will be discussed in the next section.

The resource requirements of the T3NS algorithm are presented in
table~\ref{tab:compl}. The scaling is explained in appendix~\ref{ap:complexity}.
An example of a T3NS wave function ansatz is given in fig.~\ref{fig:ansatzex}.
\begin{table}[h!] 
  \centering 
  \resizebox{\columnwidth}{!}{%
    \begin{tabular}{r c | c} & DMRG & T3NS \\
    \cline{2-3} & &\\
    CPU time: & $\order{k^4D^2 + \underline{k^3D^3}}$  & $\order{k^4D^2 + \underline{k^3D^4}}$\\
    Memory:   & $\order{k^3D^2}$                       & $\order{k^3D^2 + kD^3}$\\
  \end{tabular}}

  \caption{\label{tab:compl}Resource requirements for DMRG and T3NS for quantum
    chemistry with renormalized operators. The underlined terms correspond with
    the complexity of the most intensive part of the algorithm, i.e. the
    matrix-vector product used in the iterative solver. The number of orbitals
    (or physical tensors in the network) is denoted by $k$ and the maximal
    virtual bond dimension is denoted by $D$. 
  } 
\end{table}

\section{Symmetries in T3NS}
\label{sec:sym}
When representing the wave function in a tensor network ansatz, the encoding of
symmetries into the network can facilitate the calculations.  Depending on the
particular tensor network ansatz used, the implementation of symmetries can be
relatively straightforward or more involved.  \cite{10.1209epli2002-00393-0,
10.10881742-5468200710P10014, 10.1103PhysRevA.82.050301,
10.10881367-2630123033029, 10.1016j.cpc.2014.01.019, 10.1016j.aop.2012.07.009,
10.1103PhysRevB.86.195114, 1808.10804v2, PhysRevB.78.245109, 1408.5039v1}
The usage of symmetries in the T3NS is facilitated by restricting to a maximum
of three legs for every tensor in the wave function ansatz. This makes the
treatment of both abelian and non-abelian global symmetries very analogous to
the MPS, where this is already well studied\cite{10.1209epli2002-00393-0,
10.10881742-5468200710P10014, 10.1103PhysRevA.82.050301,
10.10881367-2630123033029, 10.1016j.cpc.2014.01.019, PhysRevB.78.245109,
1408.5039v1}.

\subsection{Labeling the basis states}
\label{sec:bstates}
In the T3NS the index of a virtual (or physical) leg, specifies the different
virtual (or physical) basis states traveling through this leg. The three-legged
tensors are, due to the Wigner-Eckart theorem, irreducible tensor operators of
the total symmetry group. The basis states need to transform according to the
rows of the irreducible representations of the symmetry
group.\cite{10.1209epli2002-00393-0, 10.10881742-5468200710P10014,
10.1103PhysRevA.82.050301, 10.10881367-2630123033029, 10.1016j.cpc.2014.01.019}
Each basis state $\ket{\alpha}$ (or index of a leg) can thus be labeled with the
irrep and the row of the irrep according to which it transforms, i.e.
$\ket{\alpha} = \ket{\text{Irreps and rows}, \alpha'}.$

The label $\alpha'$ is needed to discern the different basis states belonging to
the same list of irreps and rows of irreps. For a physical bond in
non-relativistic quantum chemistry for example we need the local basis states at
every orbital. We get
\begin{align}
  \ket{-} &= \ket{\pi=0, N_{\uparrow}=0, N_{\downarrow}=0}\\
  \ket{\uparrow} &= \ket{\pi=1, N_{\uparrow}=1, N_{\downarrow}=0}\\
  \ket{\downarrow} &= \ket{\pi=1, N_{\uparrow}=0, N_{\downarrow}=1}\\
  \ket{\uparrow\downarrow} &= \ket{\pi=0, N_{\uparrow}=1, N_{\downarrow}=1}
  \intertext{for the parity $\mathbb{Z}_2$ $(\pi)$, and two $U(1)$-symmetries
  for both the spin up and down $(N_\uparrow, N_\downarrow)$, or}
  \ket{-} &= \ket{\pi=0, I=I_0, N=0, j=0, m=0}\\
  \ket{\uparrow} &= \ket{\pi=1, I=I_k, N=1, j= \frac{1}{2}, m=\frac{1}{2}}\\
  \ket{\downarrow} &= \ket{\pi=1, I=I_k, N=1, j= \frac{1}{2}, m=-\frac{1}{2}}\\
  \ket{\uparrow \downarrow} &= \ket{\pi=0, I=I_0, N=2, j=0, m=0}
\end{align}
for $\mathbb{Z}_2$ $(\pi)$, $U(1)$ $(N)$, $SU(2)$ $(j, m)$ and the real abelian
point-group symmetries $(I)$. The labels $\pi, N_\uparrow, N_\downarrow, N, j$
and $I$ represent irreps of the different symmetries, while $m$ labels the row
of irrep $j$. $I_k$ is the point-group irrep of the orbital.  A double
occupation results in the trivial point group irrep $I_0$ since $I_k \otimes I_k
= I_0$ for real abelian point group symmetries.  In this example, $SU(2)$ is the
only symmetry that needs an extra label for the row since the other symmetries
have one-dimensional representations.

We omitted the label $\alpha'$, since the irreps and rows already uniquely
label the local physical basis states. However for the labeling of the virtual
basis states the label $\alpha'$ is still needed.

For calculations with fermions we utilize fermionic tensor networks. For every
performed permutation or contraction the fermionic signs are calculated by
looking at the parities of the different basis
states.\cite{10.1103PhysRevB.95.075108, 10.1021acs.jctc.8b00098} One could note
that labeling the parity $\pi$ is redundant since is it already fixed by the
total number of particles in the state. However, we chose to keep explicitly
track of the parity of the states. In this way, we can separate completely the
fermionic sign handling from the particle numbers in the basis states. This
allows a more modular implementation of the different symmetries and the
fermionic signs.

\subsection{Reduced tensors}
\label{sec:reducedtensor}
\begin{table*}[!ht]
  \begin{align}
    \label{eq:ansatztens}
    \mathbf{A} &= \sum_{\begin{smallmatrix}
        \fdof{1}\\\fdof{2}\\\fdof{3} \end{smallmatrix}} 
    A_{\begin{smallmatrix}\fdof{1}\\\fdof{2}\\\fdof{3}\end{smallmatrix}} 
    |\fdof{1}\rangle |\fdof{2}\rangle \langle\fdof{3}|\\
    \label{eq:WEtheorem}
  &= \sum_{\dofa{1}\dofa{2}\dofa{3}}
  \fdI{1}{2}{3} \fdz{1}{2}{3}\fdN{1}{2}{3} 
  \langle j_1 m_1 j_2 m_2 | j_3 m_3 \rangle
  \tilde{T}_{\rdofa{1}\rdofa{2}\rdofa{3}}
  |\dofa{1}\rangle |\dofa{2}\rangle \langle\dofa{3}|\\
  \label{eq:WtheoremWigner}
  &= \sum_{\dofa{1}\dofa{2}\dofa{3}} \dz\dI\dN \tj{1}{2}{3} [j_3] \msign{3}
  T_{\rdofa{1}\rdofa{2}\rdofa{3}}|\dofa{1}\rangle |\dofa{2}\rangle
  \langle\dofa{3}|.
  \end{align}
  \begin{equation}
    \label{eq:ClebschWigner}
    \langle j_1 m_1 j_2 m_2 | j_3 m_3 \rangle = \tj{1}{2}{3} \sqrt{2j_3 +1} 
    (-1)^{j_1 - j_2 + m_3}.
  \end{equation}
\end{table*}
By using the labeling discussed in section \ref{sec:bstates}, we can write every
three-legged tensor in the wave function ansatz as shown in
eq.~\ref{eq:ansatztens}. Due to the Wigner-Eckart theorem, the reducible tensor
$A$ can be rewritten as a reduced tensor $\tilde{T}$ multiplied with the
Clebsch-Gordan coefficients of the different symmetries as shown in
eq.~\ref{eq:WEtheorem}. These coefficients are Kronecker deltas for the real
abelian symmetries and are the well-known Clebsch-Gordan coefficients for the
recoupling of spins for $SU(2)$.

In eq.~\ref{eq:WtheoremWigner} the latter is replaced by the Wigner-3j symbol 
using eq.~\ref{eq:ClebschWigner}. Furthermore, several shorthand notations are 
introduced. \dof{i} is shorthand notation for the full labeling of irreps and
rows of irreps, while \rdof{i} is the labeling of only the irreps and not the
rows ($m$) of the basis state.

Shorthand notations are also introduced for the Kronecker deltas and for the
square root of the multiplicity of $j_i$ $([j_i] = \sqrt{2j_i + 1}$). A phase is
absorbed in the reduced tensor $T$ (hence the transition from $\tilde{T}$ to
$T$). We found the calculations particularly simple with this absorbed phase.

Eq.~\ref{eq:WtheoremWigner} expresses the reducible tensor $A$ in terms of  
a reduced tensor $T$ independent of the labels of the rows of irreps in \dof{i}
and a \emph{symmetry tensor}.  This symmetry tensor consists of the remaining
terms in eq.~\ref{eq:WtheoremWigner}. It contains the complete dependency of $A$
on the rows of irreps and is completely independent of the labels $\alpha_i$.

To construct the wave function, the different tensors in the network should be
contracted. The physical tensors at the border of the network have one dangling
uncontracted virtual bond since we assume that all physical tensors have two
virtual bonds.  For all but one bordering sites, this virtual bond of reduced
dimension 1 will correspond with a ket state in eq.~\ref{eq:WEtheorem} with the
same quantum numbers as the vacuum state. For the remaining bordering site, the
dangling virtual bond of reduced dimension 1 will correspond with a bra state in
eq.~\ref{eq:WEtheorem} with the same quantum numbers as the target state. By
changing the allowed quantum numbers in this bond, we can easily change the
quantum numbers of the targeted state. This is equivalent with the
singlet-embedding strategy introduced by Sharma and Chan for spin-adapted
DMRG\cite{1408.5039v1}.

Since the symmetry tensors targets all equivalent states in a multiplet at once,
the complete wave function should be normalized by dividing by
$[j_\text{final}]$, with $j_\text{final}$ the total spin quantum number of the
target state. 

The original network of reducible tensors representing the wave function (e.g.\
as given in fig.~\ref{fig:ansatzex}) now factorizes into two networks with the
same shape.  One network consists of reduced tensors. It covers the complete
dependency of the labels $\alpha_i$ and is also dependent of the labeling of
irreps.  The other network is built from symmetry tensors. The dependency of the
wave function on the labeling of the rows of irreps ($m$) is completely captured
by this network. Furthermore, this network is completely independent of the
labels $\alpha_i$.

\subsubsection{Graphical depiction}
\label{sec:graph}
\begin{figure}[!ht]
  \centering
  \includegraphics[width=\columnwidth]{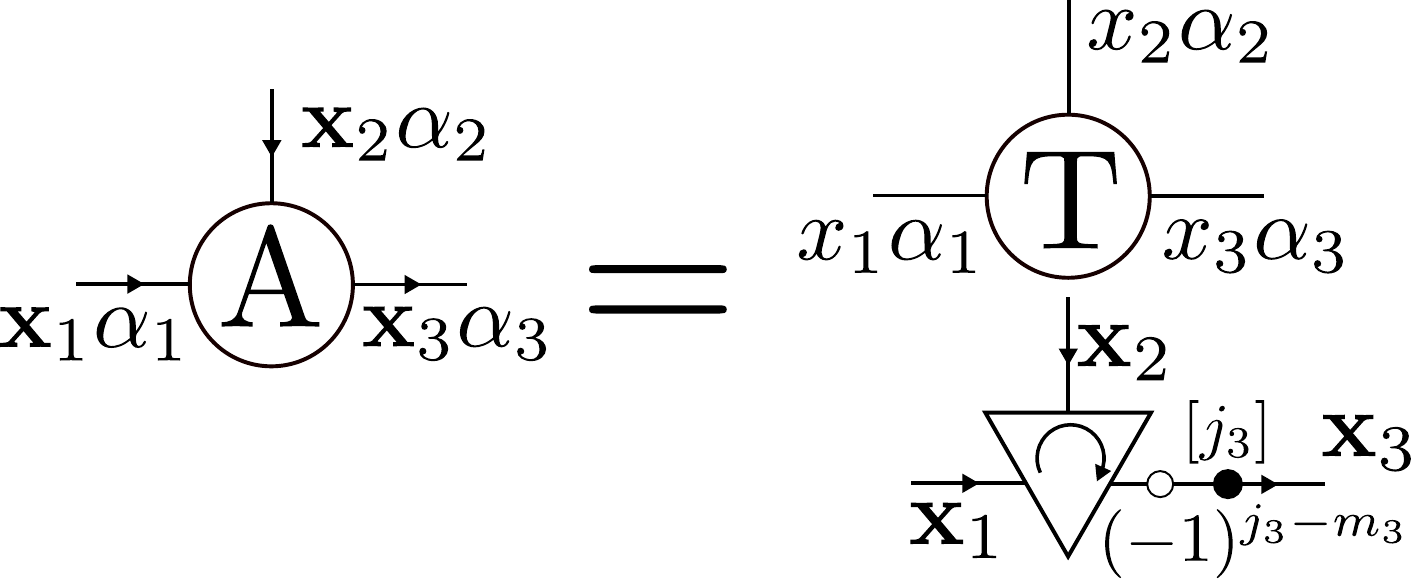}
  \caption{\label{fig:ansatz}The ansatz for a three-legged tensor with
    symmetries. This is a graphical depiction of eq.~\ref{eq:WtheoremWigner}.
    The reduced tensor and the symmetry tensor are represented by the upper and
    lower part, respectively.
  }
\end{figure}
In fig.~\ref{fig:ansatz} we introduce a graphical depiction of
eq.~\ref{eq:WtheoremWigner}. 
The triangle represents the Wigner-3j symbol and the various Kronecker deltas.
The hollow circle represents $\msign{3}$ while the full circle represents
$[j_3]$. 

The direction of the arrows on the legs fixes the sign of $m_i$ in the Wigner
sign. An incoming arrow corresponds with $m_i$, while an outgoing arrow is
$-m_i$. Such a convention is also needed for a correct treatment of the
fermionic signs.\cite{10.1021acs.jctc.8b00098, 10.1103PhysRevB.95.075108} An
incoming leg for \dof{i} represents \ket{\dof{i}} while an outgoing leg
represents \bra{\dof{i}}.

The arrow in the triangle is needed to fix the order of the bra and kets in 
eq.~\ref{eq:WtheoremWigner}. The arrow runs from the first bra or ket to the 
last bra or ket. Since the tensor is not invariant under permutation of these 
bras and kets, due to its fermionic character, the fixing of the order in our 
graphical depiction is needed.  The arrow also represents the cyclic invariance 
of the Wigner-3j symbol. A phase has to be included for the Wigner symbol when 
changing the direction of the arrow but not when rotating the arrow. A quick 
summary of the graphical depiction is also given in appendix~\ref{ap:graph}.

\subsubsection{Sparsity and compression of the tensors}
\label{sec:compr}
The symmetry tensor encodes a lot of sparsity since it consists out of several 
Kronecker deltas and a Wigner-3j symbol which respects the triangle inequality. 
In this way, it is easy to see which elements of the reduced tensor $T$ do not 
contribute to the resulting reducible tensor $A$. Such index combinations can be 
omitted from the optimization.

Next to sparsity, the symmetry tensor also compresses the data. This is clear 
since both the symmetry tensors and the reduced tensors need to be summed over 
$\rdof{i}$ (or $\pi_i I_i N_i j_i$) when contracting, while only the symmetry 
tensor need to be summed over $m_i$ for a contraction. This last summation is 
completely independent of the tensor $T$, simplifying the calculations.

\subsection{The Chemical Hamiltonian}
\label{sec:ChemHam}
The non-relativistic quantum chemical Hamiltonian is given by
\begin{equation}
  H = \sum_{ij} t_{ij} \sum_{\sigma} \cre{i\sigma} \an{j\sigma} + 
  \frac{1}{2} \sum_{ijkl}V_{ijkl} \sum_{\sigma\tau}\cre{i\sigma}\cre{j\tau}
  \an{l\tau}\an{k\sigma},
\end{equation}
where $i,j,k$ and $l$ are indices for the different spatial orbitals and
$\sigma,\tau$ represent the spin degree of freedom $(\uparrow$ or $\downarrow)$.

For the four point interactions this separates into the following cases:
\begin{align}
  \label{eq:Viikk}
  V_{iikk}&\cre{i\uparrow}\cre{i\downarrow}\an{k\downarrow}\an{k\uparrow}, 
          &i=j,k=l\\
  \label{eq:Viikl}
  V_{iikl}&\left(\cre{i\uparrow}\cre{i\downarrow}\an{l\downarrow}\an{k\uparrow}
    + \cre{i\uparrow}\cre{i\downarrow}\an{k\downarrow}\an{l\uparrow}\right),  
          &i=j,k<l\\
  \label{eq:Vijkk}
  V_{ijkk}&\left(\cre{i\uparrow}\cre{j\downarrow}\an{k\downarrow}\an{k\uparrow}
    + \cre{j\uparrow}\cre{i\downarrow}\an{k\downarrow}\an{k\uparrow}\right),  
          &i<j,k= l\\
  V_{ijkl}&(\cre{i\uparrow}\cre{j\uparrow}\an{l\uparrow}\an{k\uparrow} + 
  \cre{i\uparrow}\cre{j\downarrow}\an{l\downarrow}\an{k\uparrow} +\nonumber\\
          &\ \cre{i\downarrow}\cre{j\uparrow}\an{l\uparrow}\an{k\downarrow} +
  \cre{i\downarrow}\cre{j\downarrow}\an{l\downarrow}\an{k\downarrow}) 
          &i<j,k<l\nonumber\\
  \label{eq:Vijkl}
  + &\left(k \leftrightarrow l\right),
\end{align}
where we only used $V_{ijkl} = V_{jilk}$. The complete Hamiltonian can be
constructed by summing these terms for all possible $(i,j,k,l)$-combinations.

The creation and annihilation operators \cre{k\sigma} and \an{k\sigma} do not
yet transform according to the rows of the $SU(2)$-irreps. As is well
known\cite{10.1016j.cpc.2014.01.019}, an additional phase has to be introduced.
One possible transformation is given by
\begin{align}
  \tilde{c}^\dagger_{k\sigma} &= \cre{k\sigma}\\
  \tilde{c}_{k\sigma} &= (-1)^{\frac{1}{2} + \sigma} \an{k-\sigma}.
\end{align}
In this way, we can again split off the Clebsch-Gordan coefficients into a
symmetry tensor.
\begin{figure}[!ht]
  \centering
  \includegraphics[width=0.8\columnwidth]{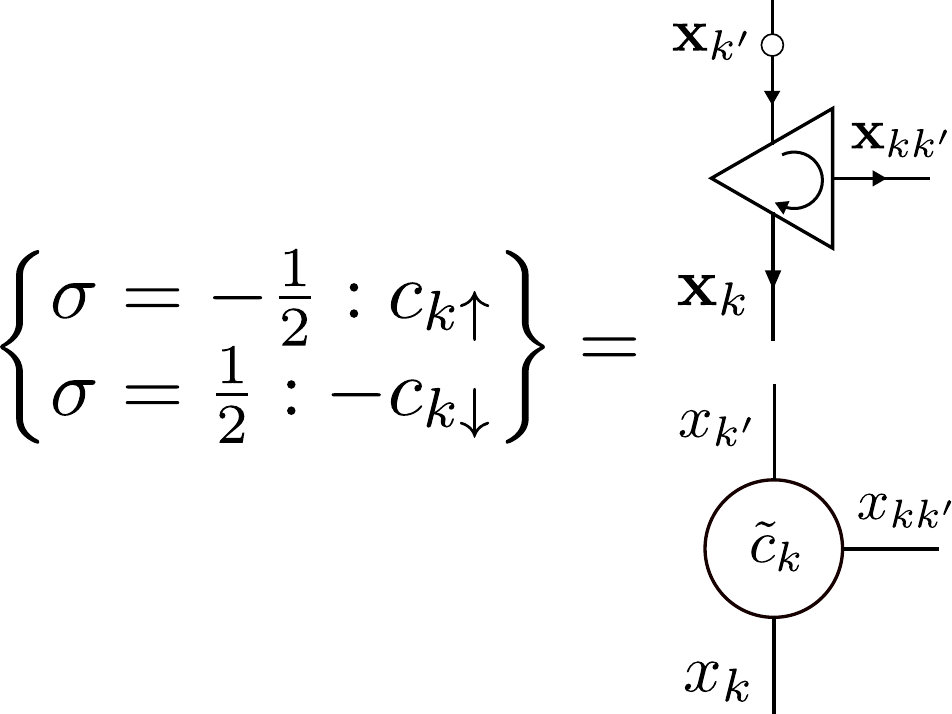}
  \caption{Graphical depiction of the annihilation operator on site $k$. The
    operators are represented by a symmetry tensor and a reduced tensor. \dof{k}
    and \dof{k'} are the local basis states for orbital $k$. The third index
    $\dof{kk'} = \left(1, I_k, -1, \frac{1}{2}, \sigma \right)$ serves the
    purpose of correctly coupling different operators. No bonds here need an
    extra $\alpha$ label, as stated in section~\ref{sec:bstates}.
  }
  \label{fig:operators}
\end{figure}
The decomposition of the annihilation operators into a reduced tensor and a
symmetry tensor is graphically depicted in fig.~\ref{fig:operators}. One can
easily calculate the tensor elements for the reduced tensors in
fig.~\ref{fig:operators}. For the creation operators the graphical depiction is
completely equivalent.

\begin{figure*}[!ht]
  \centering
  \includegraphics[width=0.9\textwidth]{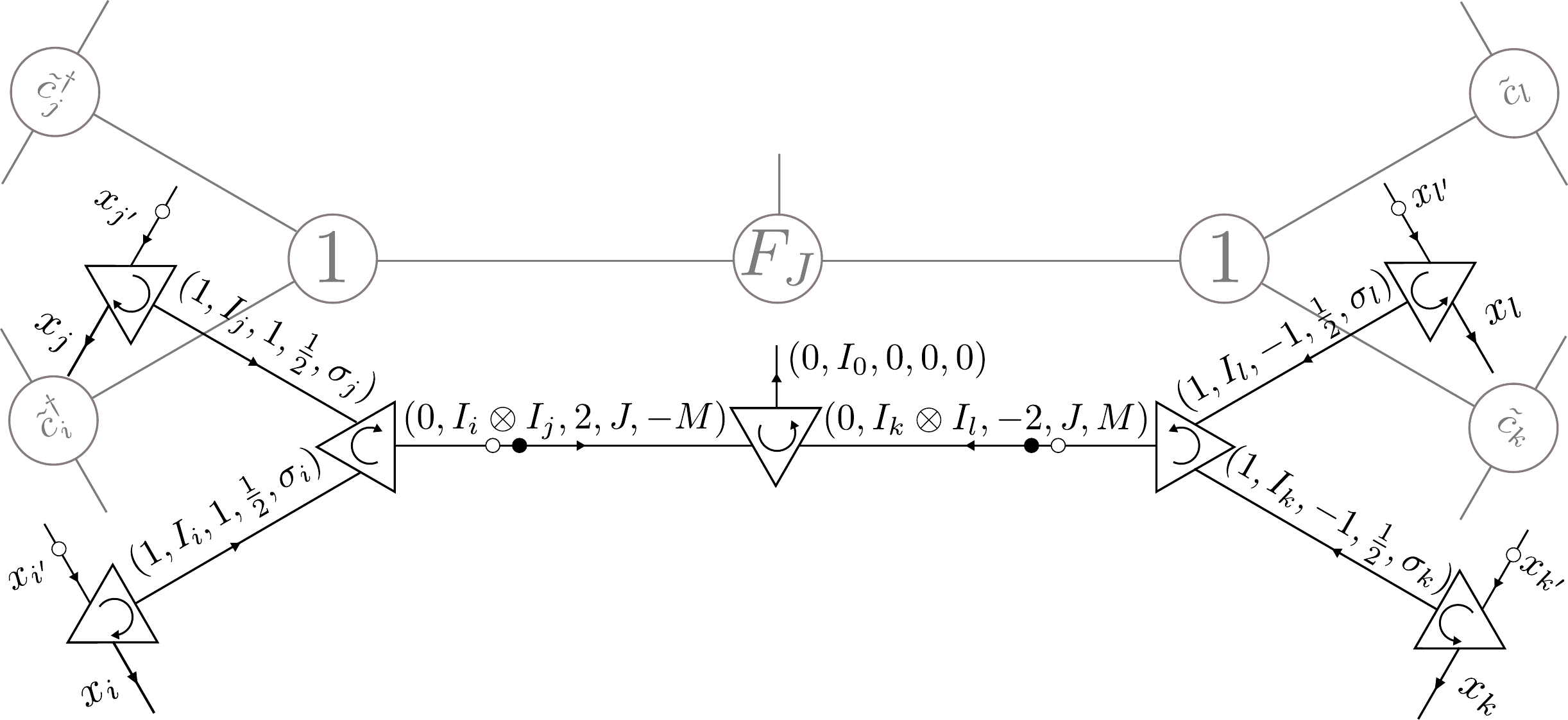}
  \caption{\label{fig:Vijkl}Graphical depiction of the terms in
    eq.~\ref{eq:Vijkl}. The upper gray layer are the reduced tensors and the
    lower layer are the symmetry tensors. Labels of connected bonds are written
    fully for clarity. The different operators need to couple to the trivial
    singlet, since the term in eq.~\ref{eq:Vijkl} transforms trivially under the
    used symmetries.}
\end{figure*}

The most illustrative example is given by eq.~\ref{eq:Vijkl} since here four
different spatial orbitals can be involved. The construction of these terms is
graphically shown in fig.~\ref{fig:Vijkl} where
\begin{equation}
  F_J = - [J]\left(V_{ijkl} + (-1)^J V_{ijlk}\right).
\end{equation}

The tensor network in fig.~\ref{fig:Vijkl} should be manipulated to the same
geometry of the tensor network of the wave function. For obtaining a T3NS
structure out of fig.~\ref{fig:Vijkl}, two types of transformations are needed.
First, the order of the operators (e.g.\ switching $i$ and $k$) can be changed
by using properties of the Wigner symbols and taking an appropriate fermionic
sign into account.  One can also insert identities into the network in
fig.~\ref{fig:Vijkl}. These two transformations suffice to change the network in
fig.~\ref{fig:Vijkl} to an arbitrary T3NS network.

All terms in 
eq.~\ref{eq:Viikk},~\ref{eq:Viikl},~\ref{eq:Vijkk}~and~\ref{eq:Vijkl} 
can be manipulated in this way. The most important point here is that both the
wave function ansatz and the Hamiltonian are represented by the same tensor
network shape and both are factorized into a reduced tensor network and a
symmetry tensor network.

\subsection{The optimization}
\label{sec:optimization}
In the previous sections~\ref{sec:reducedtensor} and \ref{sec:ChemHam}, we have
shown that both the wave function ansatz and the Hamiltonian can be factorized
into a reduced tensor network and a symmetry tensor network.
In this section, we briefly discuss how this representation can be used during
the optimization algorithm of the T3NS.

The optimization of a T3NS occurs in a similar way as for DMRG, i.e.\ we sweep
through the network and optimize only two tensors at a time. During this local
optimization of the network, the effect of the Hamiltonian on the other tensors
(i.e.\ the environment) can be efficiently captured by renormalized
operators.\cite{10.10631.478295, 10.10631.1449459, 10.1016j.cpc.2014.01.019,
10.10631.3152576} The usage of renormalized operators reduces the quartic
scaling of the total chemical Hamiltonian as a function of the number of
orbitals to a quadratic scaling for the effective Hamiltonian.

\begin{figure}[!ht]
  \centering
  \includegraphics[width=0.4\columnwidth]{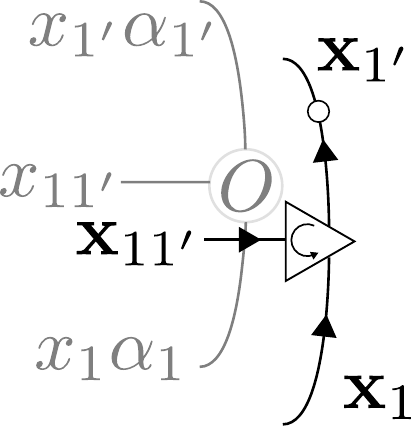}
  \caption{\label{fig:rightrenormops}Graphical depiction of a renormalized
    operator. Both the symmetry tensor and the reduced tensor $O$ are shown.
    The index \dof{1} (\dof{1'}) originates from the T3NS ansatz of the ket
    (bra) wave function. The index \dof{11'} corresponds with a bond of the
    network representation of the Hamiltonian. The tensor $O$ is the resulting
    reduced tensor from partial contraction of the triple-layered network.
  }
\end{figure}

These renormalized operators are partial contractions of the Hamiltonian
and the T3NS wave function. More specifically, the energy of a wave function
$E~=~\langle \Psi | H | \Psi \rangle$ can be calculated by sandwiching the
Hamiltonian terms in their network form between the wave function and its
adjoint in the same network form. This triple-layered network of reduced and
symmetry tensors can be fully contracted to obtain the energy. However, during
the optimization of only two sites of the network, a lot of the triple-layered
network can be precontracted and reused since it does not change during this
optimization step. The renormalized operators are exactly these precontracted
parts of the network. A graphical depiction of a renormalized operator is
given in fig.~\ref{fig:rightrenormops}. This is an object which has indices
\dof{1}, \dof{1'} and \dof{11'}. These indices are explained in the caption of
fig.~\ref{fig:rightrenormops}.

After the optimization of the two sites, the sweep algorithm moves on to a
neighboring pair of two sites, which has one site in common with the previous
pair. These sites are optimized by a new effective Hamiltonian. In order to
calculate this effective Hamiltonian through renormalized operators, one can
recycle certain of these operators and update others. This is in fact equivalent
with the usage of renormalized operators in DMRG for quantum chemistry. 

\begin{figure*}[!ht]
  \centering
  \includegraphics[width=0.9\textwidth]{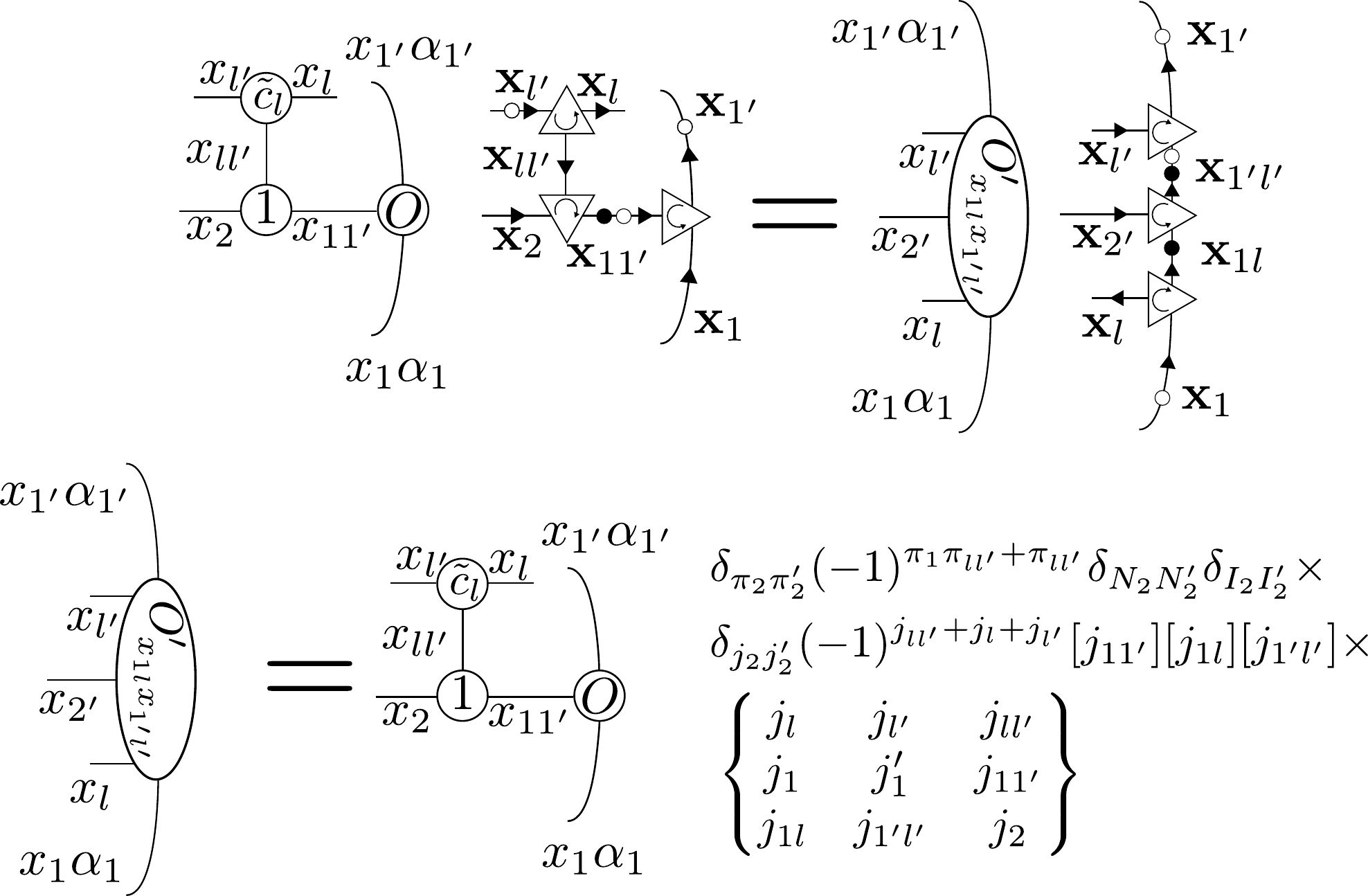}
  \caption{\label{fig:siteappend}The appending of a site-operator $\tilde{c}_l$
    to a renormalized operator is shown. The symmetry tensors change of
    form to ease following calculations in the algorithm. The reduced tensors 
    are contracted and the result is permuted appropriately. Only index $1$ and
    $1'$ need an extra $\alpha$-label to discern between the different basis
    states.
  }
\end{figure*}
In fig.~\ref{fig:siteappend}, an example of such an update is given. A site
operator (as given in fig.~\ref{fig:operators}) is appended to an already
existing renormalized operator. The resulting tensor is then manipulated to
another form, for an easier construction of the effective Hamiltonian. In the
example, the manipulation of the symmetry tensors gives rise to extra factors. A
Wigner-9j symbol arises due to the recoupling of the spin basis states. The
phase $(-1)^{\pi_1 \pi_{ll'} + \pi_{ll'}}$ originates from the fermionic
character of the tensors.

Wigner symbols are used for the update of renormalized operators and the
construction of the effective Hamiltonian when a branching tensor is optimized.
However, when only physical tensors are optimized, no Wigner symbols are needed.
This is discussed in more depth in appendix~\ref{ap:optimize}.
The optimization of two adjacent physical tensors is exactly the same as in
DMRG. 

As a side remark, the present implementation, omitting branching tensors, can
perform a DMRG algorithm which only needs Wigner symbols during the update of
renormalized operators and not during the iterative optimization step.

More details for the construction of the effective Hamiltonian and the updating
of the renormalized operators are given in appendix~\ref{ap:Arithmetics}.

\section{Calculations}
\label{sec:calc}
In this section, we present several calculations with T3NS. The used
implementation for the calculations can be found on github\cite{T3NSsourcecode}.
This implementation is able to exploit $\mathbb{Z}_2$, $U(1)$, $SU(2)$ and the
real abelian point group symmetries. The symmetries can be included in a modular
way.  This enables us to compare calculations with and without $SU(2)$ and point
group symmetries for the same implementation.

After the optimization of two sites, this two-site object has to be split into
two separate sites again. This is done by the Singular Value Decomposition
(SVD). The truncation of the virtual bond dimension during this step is done in
two different ways.

First, a fixed maximal bond dimension can be imposed. The algorithm will keep as
many singular values into consideration as possible. It will first select the
largest singular values until no non-zero singular values are left or the
maximal bond dimension is reached. The remaining singular values and their
corresponding basis states are discarded.

Second, the dynamic block state selection (DBSS) can be
used\cite{10.1103PhysRevA.83.012508, doi:10.1002qua.24898}. With this method,
the algorithm keeps the largest singular values until a threshold for a cost
function is reached. The cost function used in our implementation is given by
\begin{equation}
  w_{disc} = \sum_{i_{disc}} s_{i_{disc}}^2
\end{equation}
i.e.\ the sum of the squares of all discarded singular values. This corresponds
with $\langle \Psi_{disc} | \Psi_{disc} \rangle$, with $|\Psi_{disc}\rangle$ the
discarded part of the wave function during truncation. Other cost functions can
be easily implemented. 

Next to a targeted threshold, a minimal and maximal bond dimension should be
specified. Throwing away too many basis states at a certain stage can impede the
optimization at later stages, even though the truncation error is only minimal
at that point. Specifying a minimal bond dimension ensures a certain flexibility
at all time. The maximal bond dimension is needed to prevent a large increase in
both run time and memory usage when the imposed threshold can not be reached.

As noted in section~\ref{sec:compr}, the usage of symmetries with irreps that
are more than one-dimensional, such as $SU(2)$, introduces a compression of the
wave function. Different basis states belonging to the same multiplet can be
represented by a singular reduced basis state. Analogous to the bond dimension
being the number of basis states kept in the bond, the \emph{reduced bond
dimension} is defined as the number of reduced basis states kept. When using
$SU(2)$, the reduced bond dimension of the bonds, and not the bond dimension,
will reflect the computational complexity.

Just as with DMRG for quantum chemistry, keeping track of the renormalized
operators is the most taxing part on memory for T3NS. The amount of 
renormalized operators needing to be stored for T3NS are of the same order as
for DMRG. However, T3NS calculations can be performed on a considerably lower
bond dimension for a similar accuracy. Consequently, this lowers the storage
requirements for the renormalized operators and allows us to keep all tensors on
memory at all time. No checkpoint files need to be written to disk or read from
disk during the algorithm for the present system sizes and bond dimensions.

\subsection{The Bisoxo and Peroxo Isomer of \cuo}
We revisit the bis($\mu$-oxo) and the $\mu=\eta^2:\eta^2$ peroxo \cuo isomers as
a test case for T3NS with $SU(2)$ and abelian point group symmetries.  These
transition-metal clusters have been previously studied by other ab initio
methods such as the complete active space self-consistent field theory (CASSCF),
the complete active space self-consistent field theory with second order
perturbation theory (CASPT2),\cite{10.1021jp056791e} the restricted active space
self-consistent field theory with second order perturbation theory
(RASPT2)\cite{10.10631.2920188}, DMRG\cite{10.1103PhysRevA.83.012508,
doi:10.1021acs.jctc.6b00714} and DMRG+CT\cite{doi:10.10631.3275806} (DMRG with
canonical transformation theory).  It was the largest system studied in the
initial T3NS paper using only $U(1)\otimes U(1)$
symmetry\cite{10.1021acs.jctc.8b00098} (i.e.  conservation of both particle
number and spin projection).

We perform calculations for both isomers in an (26e, 44o) active space.  The
same active space is used as in refs.~\citenum{10.1021acs.jctc.8b00098,
10.1103PhysRevA.83.012508}. Both isomers have a $D_{2h}$ point group symmetry
and their ground state is a singlet state in the $A_g$ irrep of
$D_{2h}$.\cite{10.1103PhysRevA.83.012508} When using the $SU(2)$ and/or point
group symmetry adapted version of T3NS, states corresponding to these irreps
will be targeted.  The same T3NS shape and orbital ordering is used as in
ref.~\citenum{10.1021acs.jctc.8b00098}. We also perform calculations of the
lowest lying triplet state in the $A_g$ irrep of $D_{2h}$ for both isomers.

\subsubsection{The Bisoxo isomer with and without spin symmetry}
In order to compare the present spin adapted version of T3NS with its 
non-adapted predecessor\cite{10.1021acs.jctc.8b00098}, we perform several 
calculations for the bisoxo isomer at different bond dimensions and with
different symmetries included. 

\begin{figure*}[!ht]
  \centering
  \includegraphics[width=\textwidth]{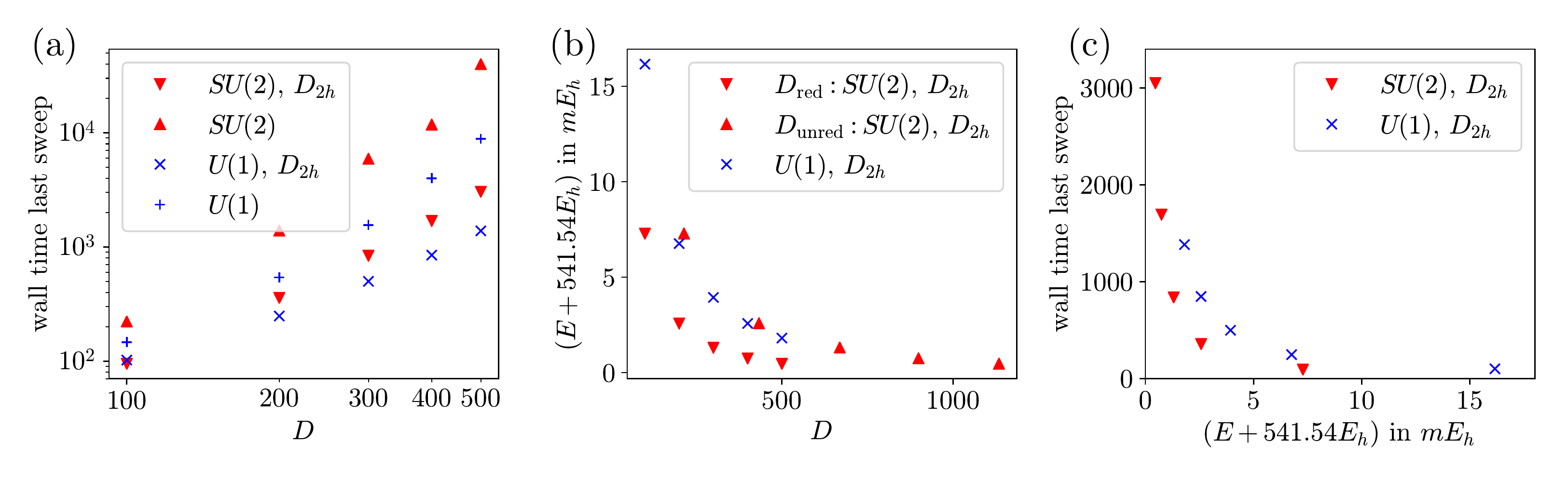}
  \caption{Results for the bisoxo isomer obtained on an 18-core Intel Xeon Gold
    6140 (Skylake at 2.3 GHz). Calculations with a fixed maximal bond dimension
    of $D=100,200,300,400$ and $500$ were performed.  Results for $U(1) \otimes
    U(1)$ combined with or without $D_{2h}$ are given by crosses.  Results for
    $U(1) \otimes SU(2)$ combined with or without $D_{2h}$ are given by
    triangles. The bond dimensions are in this case the reduced bond dimensions,
    except for (b) where also unreduced bond dimensions are shown. (a) shows the
    wall time of the last sweep for different bond dimensions and different
    symmetries. (b) shows the ground state energy for different bond dimensions.
    For the spin-adapted calculations, both the maximal reduced as unreduced
    bond dimensions are given. (c) shows wall time of last sweep in function of
    ground state energy.
  }
  \label{fig:Cu2O2plots}
\end{figure*}
In fig.~\ref{fig:Cu2O2plots}(a), timings for the last sweep are shown for
several fixed bond dimensions. Calculations were performed with $U(1) \otimes
U(1)$ symmetry and with $SU(2) \otimes U(1)$ symmetry. For both, calculations
with and without the $D_{2h}$ point group symmetry are done. For the
spin-adapted versions, the bond dimensions shown are the reduced bond
dimensions. 

As expected, the usage of the point group symmetry introduces a lot of sparsity
in the tensors which speeds up calculations considerably. Calculation time
improved by a factor of 6 and 13  at $D=500$ for $U(1) \otimes U(1)$ and $SU(2)
\otimes U(1)$ respectively when including $D_{2h}$. For both calculations with
or without spin symmetry, the inclusion of $D_{2h}$ yields practically the same
energies and maximal truncation errors as when performing the calculation
without the point group symmetry.

For tensors of the same size, calculations including spin symmetry are
computationally more intensive than without spin symmetry as can be seen in
fig.~\ref{fig:Cu2O2plots}(a).
This is expected since the reduced tensors are more
dense than the reducible tensors. However, the compressed nature of the reduced
tensors (see section \ref{sec:compr}) makes spin-adapted calculations at a
certain \textbf{reduced} bond dimension more accurate than calculations without
spin symmetry at an equal bond dimension. This can be seen in
figure~\ref{fig:Cu2O2plots}(b). The maximal unreduced bond dimension during
$SU(2)$ calculations is also given in this figure. For the present calculations, 
the maximal unreduced bond dimension is approximately twice as large as the 
imposed maximal reduced bond dimension. When comparing wall time with achieved
accuracy, the calculations with $SU(2)$ included are considerably faster, as is 
shown in fig.~\ref{fig:Cu2O2plots}(c).

\subsubsection{The bisoxo and peroxo isomers with spin symmetry}
\begin{figure}[!ht]
  \centering
  \includegraphics[width=\columnwidth]{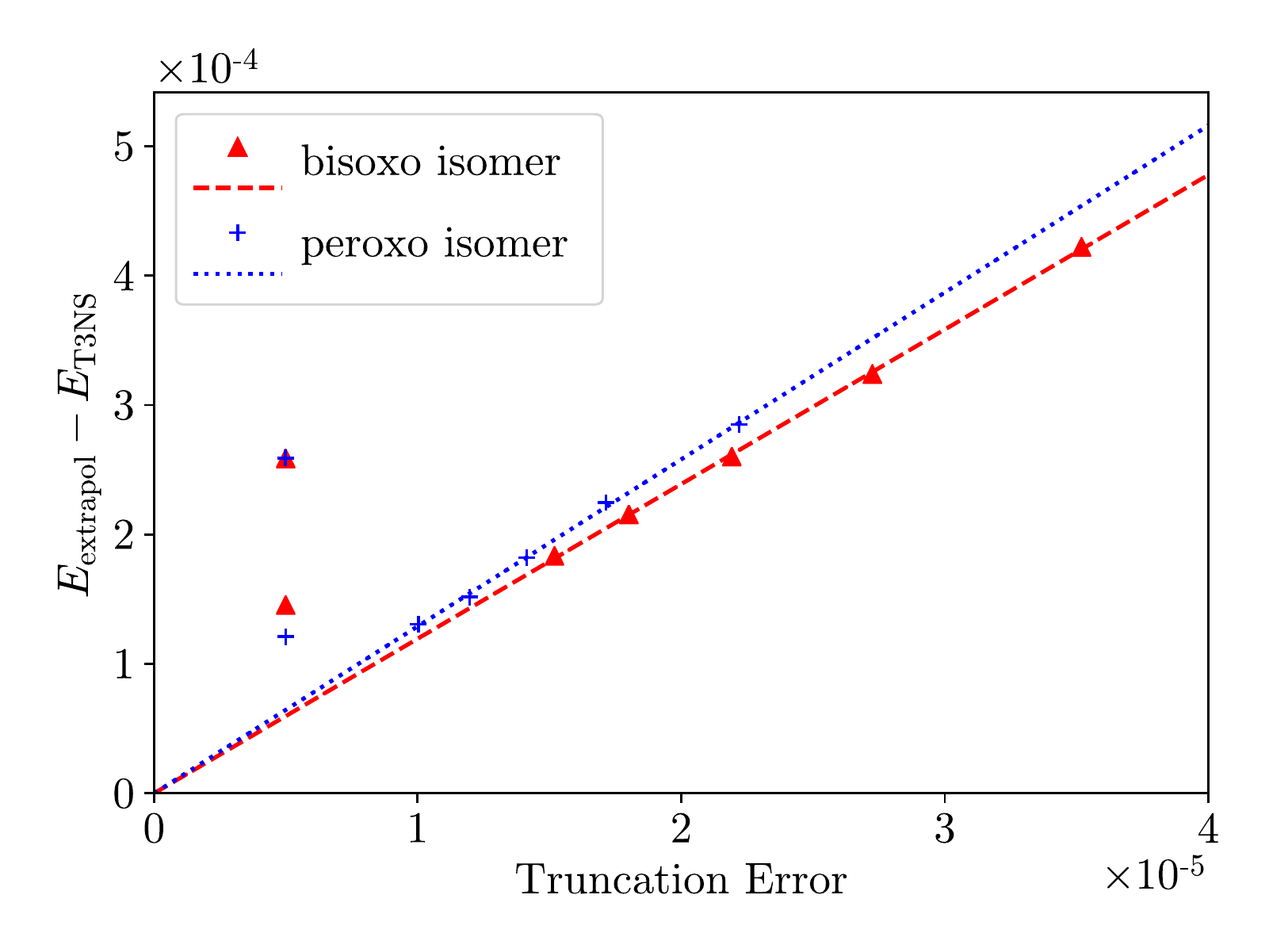}
  \caption{Extrapolation of the energy for both the bisoxo and peroxo isomer.
    Extrapolation is done by using the results for $D=600,700,800,900$ and
    $1000$.  The DBSS calculations of table~\ref{tab:Cu2O2} targeting a
    truncation error of $5\cdot 10^{-6}$ are also shown in the figure. They are
    however not used for the extrapolation.
  }
  \label{fig:epol}
\end{figure}
\begin{table}[h!] 
  \centering 
  \resizebox{\columnwidth}{!}{%
  \begin{tabular}{ l c c r }
    \hline\hline
    Method & $E_\mathrm{bisoxo}[\mathrm{E_h}]$ & $E_\mathrm{peroxo}[\mathrm{E_h}]$ & $\Delta E$
    [kcal/mol]\\
    \hline
    DMRG\cite{10.1103PhysRevA.83.012508}& -541.53853 & -541.58114 & 26.7\\
    \hline
    \multicolumn{4}{l}{T3NS with $U(1) \otimes U(1)$\cite{10.1021acs.jctc.8b00098}}\\
    500 & -541.53820 & -541.58094 & 26.8\\
    \hline
    \multicolumn{4}{l}{T3NS with $U(1) \otimes SU(2) \otimes D_{2h}$}\\
    300 & -541.53869 & -541.58119 & 26.7\\
    500 & -541.53954 & -541.58171 & 26.5\\
    $500,5 \cdot 10^{-6}$ & -541.53986 & -541.58183 & 26.3\\
    1000 & -541.53993 & -541.58197 & 26.4\\
    $1000,5 \cdot 10^{-6}$ & -541.53997 & -541.58198 & 26.4\\
    Extrapolated & -541.54012 & -541.58210 & 26.3\\
    \hline\hline
  \end{tabular}}
  \caption{Energy gaps and ground state energies between the bisoxo and peroxo 
    isomers. The energy gaps are given in kcal/mol. Ground state energies are
    given in Hartree. For T3NS calculations with fixed bond dimension, the bond
    dimension is given in the first column. For T3NS calculations using DBSS,
    the minimal bond dimension and the truncation error is given. For T3NS with
    $SU(2)$ the given bond dimensions are the reduced ones. Maximum bond
    dimensions around 2000 were reported for both clusters during the DMRG
    calculations of ref.~\citenum{10.1103PhysRevA.83.012508}.
  }
  \label{tab:Cu2O2}
\end{table}
Several calculations are performed for both isomers. Both $SU(2)$ and $D_{2h}$
symmetry are used. The inclusion of spin and point group symmetry considerably
improves our calculations and allows us to go to much larger bond dimensions.
Both a constant maximal bond dimension and DBSS are used. Some obtained results
are shown in table~\ref{tab:Cu2O2} alongside previously published results.
Calculations at a fixed reduced bond dimension of 300 already surpassed the most
accurate DMRG calculations of ref.~\citenum{10.1103PhysRevA.83.012508} and the
most accurate T3NS calculations of ref.~\citenum{10.1021acs.jctc.8b00098}.
For the most accurate calculation the maximal reduced bond dimension needed were
1626 and 1329 for the bisoxo and peroxo isomer respectively.

A linear extrapolation\cite{10.1103PhysRevB.67.125114, 10.10631.1449459} between
the truncation error and the energy is performed.  The extrapolation is based on
results with a fixed maximal reduced bond dimension of $D=600, 700, 800, 900$
and $1000$. This extrapolation is shown in fig.~\ref{fig:epol}. The results for
the DBSS calculations in table~\ref{tab:Cu2O2} are also given in the figure.
For both systems, two DBSS calculations are performed, both targeting a
truncation error of $5\cdot 10^{-6}$. They use however another minimal bond
dimension. Although the same truncation error is targeted for both DBSS
calculations, the accuracy of the two calculations is quite different due to
their different minimal bond dimensions. Because of this, only the calculations
with fixed bond dimension are used for the extrapolation. A linear extrapolation
looks justified for these calculations. The extrapolated values are given in
table~\ref{tab:Cu2O2}.

The lowest lying triplet states in the $A_g$ irrep are also calculated for both 
isomers. These states are easily targeted by changing the allowed quantum 
numbers in the outgoing target state of the T3NS (as discussed in 
section~\ref{sec:reducedtensor}) and are similar in computation time as for the 
singlet states. We obtained an energy of $E=-541.46194E_h$ and 
$E=-541.40184E_h$ for the bisoxo and peroxo isomer respectively when a maximal 
reduced bond dimension of 1000 was chosen.

\section{Conclusion}
\label{sec:conclusion}
In this paper we extended the T3NS ansatz with $SU(2)$ and real abelian point
group symmetries. This ansatz combines the computational efficiency of DMRG and
its ease of implementing symmetries with the rich entanglement representation of
the TTNS. We show that implementing spin and real abelian point group symmetries
is not much more involved than for DMRG.\cite{10.1209epli2002-00393-0,
10.10881742-5468200710P10014, 10.1103PhysRevA.82.050301,
10.10881367-2630123033029, 10.1016j.cpc.2014.01.019}

Several calculations for the bisoxo and peroxo isomers of \cuo are presented.
Calculations are performed both with and without spin and point group
symmetries. They illustrate the substantial advantages of using the $SU(2)$ and
point group symmetries of the chemical Hamiltonian. For a given accuracy, 
the computational time decreases with every included symmetry.

For these calculations, the same, rather intuitive, orbital ordering as in the
preliminary T3NS-paper is used\cite{10.1021acs.jctc.8b00098}. Advanced
techniques for the ordering of orbitals, like the usage of entanglement 
measures,\cite{10.1103PhysRevA.83.012508} and its effect on the accuracy will 
be of interest in subsequent research. Furthermore, results can be improved 
through the development of post-T3NS methods, in similarity to post-DMRG 
methods.  Some notable examples of post-DMRG methods which can readily be 
adapted are DMRG-SCF,\cite{10.10631.2883981, doi:10.10631.4885815, 
doi:10.10631.2883976} DMRG-CASPT2 \cite{10.10631.3629454, doi:10.10631.4959817} 
and DMRG-TCCSD (DMRG-tailored coupled cluster with single and double 
excitations)\cite{10.1021acs.jpclett.6b01908, 1809.07732v1}.

From an intuitive point of view, the T3NS ansatz looks a more natural 
representation of the entanglement structure of molecules than the linear MPS.
We expect that the T3NS will, in general, be able to provide a more compact and 
accurate parametrization of the wave function. If this also results in more 
efficient computations is at this moment not clear yet. However, in our 
preliminary paper\cite{10.1021acs.jctc.8b00098}, we noted that, for the few 
tested systems, T3NS needed an increasingly smaller bond dimension compared to 
the MPS with increasing system size. This supports the idea that with large 
enough system sizes the T3NS will become the tensor network of choice. To assess 
this trend, we need to study larger system sizes. The entanglement structure and 
efficiency for both the MPS and T3NS for larger system sizes will be one of the 
main focus points in subsequent research.

\begin{acknowledgement}
K.G.\ acknowledges support from the Research Foundation Flanders (FWO
Vlaanderen). Computational resources (Stevin Supercomputer Infrastructure) and
services were provided by Ghent University. We gratefully acknowledge fruitful
discussions with \"Ors Legeza and Sebastian Wouters.

\end{acknowledgement}

\appendix
\section{Computational complexity}
\label{ap:complexity}
Two differences can be noted between the computational complexity given in the
original T3NS-paper\cite{10.1021acs.jctc.8b00098} and the one given in this
paper.

First, no disk resource requirements are given here, since our current
implementation does not store intermediate results on disk. All tensors are kept
in memory at all time. 

Second, the CPU time of the T3NS has been lowered from $\mathcal{O}(k^5D^2 +
k^3D^4)$ to $\mathcal{O}(k^4D^2 + k^3D^4)$. In the previous
paper, we stated the complexity for updating renormalized operators with
branching tensors to be $\mathcal{O}(k^5D^2)$ for quantum chemistry. It was
noted that the recombination of two single operators in both sets of
renormalized operators to a complimentary double operator was the most intensive
part. The worst case scenario for this is indeed $\mathcal{O}(k^4D^2)$ and since
there are $\mathcal{O}(k)$ branching tensors in the network where this can
occur, we previously obtained $\mathcal{O}(k^5D^2)$. However not all branching
tensors will result in this worst case scaling of $\mathcal{O}(k^4D^2)$, and
only the most central ones in the network will have this scaling. A more
rigorous analysis showed an overall upper bound of $\mathcal{O}(k^4D^2)$
instead.

\section{Working with symmetry tensors}
\label{ap:Arithmetics}
In this appendix we explain the T3NS with fermionic reduced tensors and their
symmetry tensors in more depth. We begin with a short summary of the graphical 
notations used (Appendix~\ref{ap:graph}). In Appendix~\ref{ap:canonical}, we use 
the gauge freedom of the tensor network to define a canonical form of the 
tensors, in correspondence with the canonical form in DMRG. In 
appendix~\ref{ap:optimize}, we give a few examples of contractions of 
renormalized operators with two-site tensors. In appendix~\ref{ap:update}, some 
cases for updating the renormalized operators are given. 

\subsection{Shorthand and graphical notation}
\label{ap:graph}
We introduced a graphical depiction for the symmetry and reduced tensors used 
in the T3NS algorithm in section~\ref{sec:graph}. We shortly summarize the 
different used notations in table~\ref{tab:shorthand}. It is useful to notice 
the effect of changing the direction of an arrow on the Clebsch Gordan 
coefficients of the symmetries.
\begin{table}[!ht]{
    \begin{tabular}{r | c}
      \dof{i}  & \fdof{i}\\
      \hline
      \rdof{i} & \frdof{i}\\
      \hline
      \begin{minipage}{2cm}\includegraphics[width=2cm]{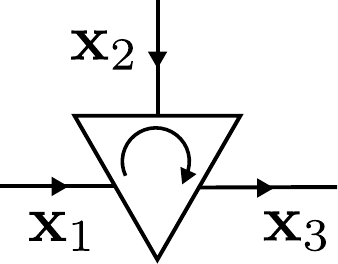}\end{minipage}
               & $\begin{array}{c}
                 \fdI{1}{2}{3} \fdz{1}{2}{3}\fdN{1}{2}{3}\\
                 \tj{1}{2}{3} \ket{\dof{1}}\ket{\dof{2}}\bra{\dof{3}}
               \end{array}$\\
      \hline
      \begin{minipage}{2cm}\includegraphics[width=2cm]{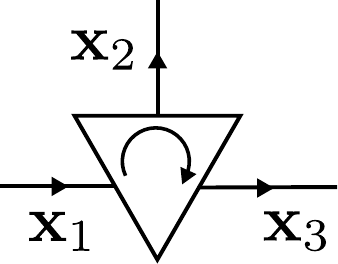}\end{minipage}
                & $\begin{array}{c}
                  \fdI{2}{3}{1} \fdz{2}{3}{1}\fdN{2}{3}{1} \\
                  \begin{pmatrix}
                    j_1 & j_2 & j_3 \\
                    m_1 & -m_2 & -m_3
                  \end{pmatrix}
                \ket{\dof{1}}\bra{\dof{2}}\bra{\dof{3}}
              \end{array}$\\
      \hline
      \begin{minipage}{2cm}\includegraphics[width=2cm]{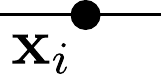}\end{minipage}
                & $\sqrt{2j_i + 1}$\\
      \hline
      \begin{minipage}{2cm}\includegraphics[width=2cm]{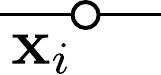}\end{minipage}
                & $\msign{i}$
    \end{tabular}}
    \caption{Used graphical depictions and shorthand notations.}
    \label{tab:shorthand}
\end{table}

\subsection{The canonical form}
\label{ap:canonical}
The representation of a wave function by a tensor network is not unique. To show
this gauge freedom, we contract the tensor $A[x]$ with an invertible matrix $Y$
and the neighboring tensor $A[x+1]$ with $Y^{-1}$\cite{10.1016j.cpc.2014.01.019,
10.1103PhysRevB.98.085155}:
\begin{align}
  \tilde{A}[x]_{ijk'} &= \sum_k A[x]_{ijk} Y_{kk'}\\
  \tilde{A}[x+1]_{k'lm} &= \sum_k (Y^{-1})_{k'k}A[x+1]_{klm}.
\end{align}
When contracting these two new tensors $\tilde{A}[x]$ and $\tilde{A}[x+1]$, we
obtain the exact result as contracting the original tensors. This freedom is
present at every virtual bond in the tensor network.

Although one can use the gauge freedom in general tensor networks to define a
canonical form,\cite{10.1103PhysRevB.98.085155, 1902.05100v1, 1903.03843v2} it
is more straightforward in loopfree finite tensor networks. In this canonical
form one tensor is chosen as orthogonality center. The currently optimized
tensor is normally chosen for this. Other tensors are orthogonal with respect to
contraction over all bonds but the one leading to the orthogonality center.
Calculating the overlap of the tensor network with itself now simplifies to a
complete contraction of the orthogonality center with its adjoint since the
contributions of the other tensors simplify to unit tensors. This allows us to
optimize the orthogonality center through an ordinary eigenvalue problem instead
of a general one.

All tensors of the T3NS wave function are of the form given in
eq.~\ref{eq:ansatztens}. Depending on the indices that are contracted, tensors
of this form can be orthogonal in different ways.

For such a reducible tensor which is orthogonal to a contraction over index 1 and
2 we propose that the reduced tensor should be orthogonal as well with respect
to a contraction over index 1 and 2.  Furthermore, the adjoint of the reduced
tensor (i.e. the corresponding reduced tensor for the bra wave function) is
given by its hermitian.
\begin{figure}[!ht]
  \centering
  \includegraphics[width=\columnwidth]{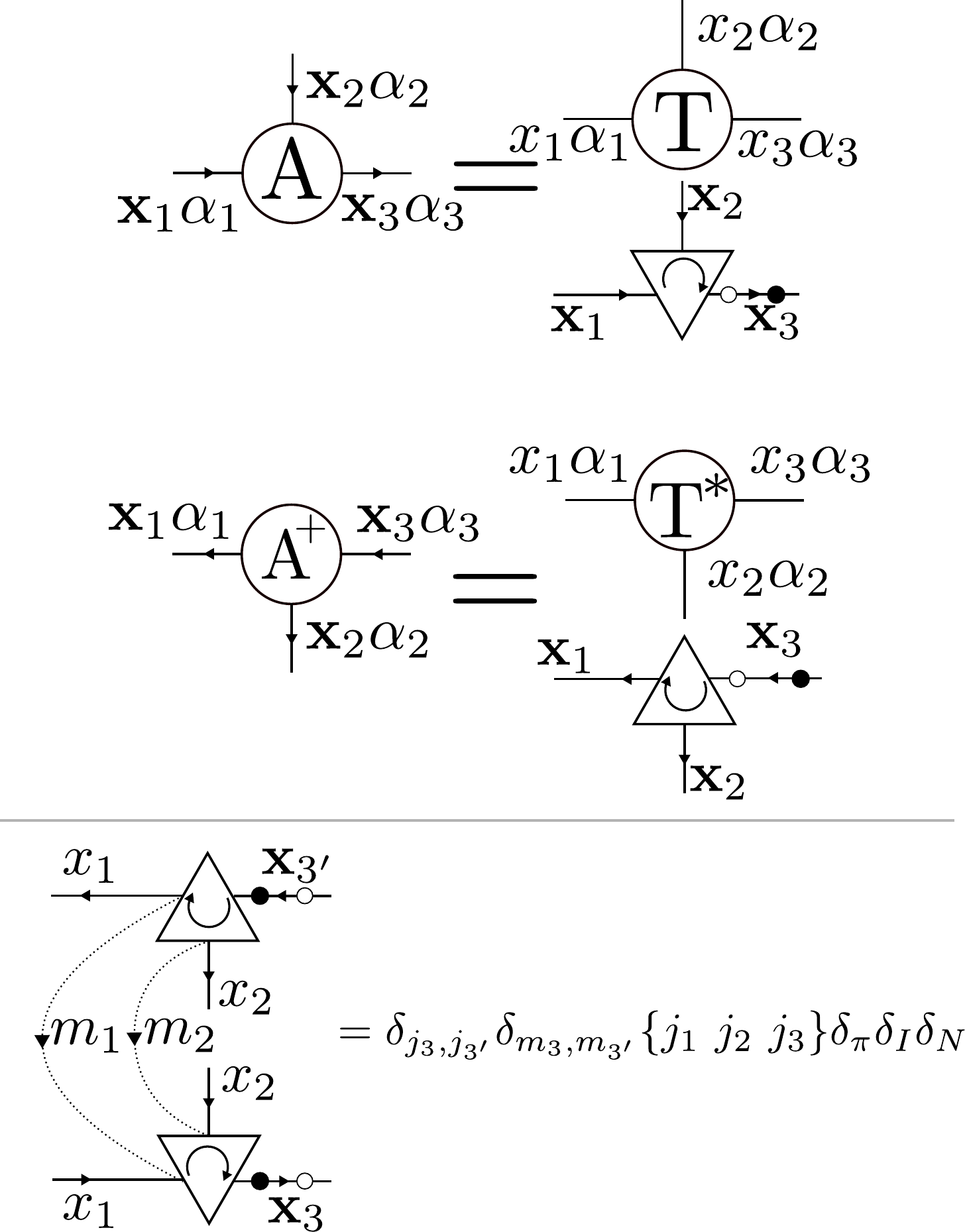}
  \caption{\label{fig:leftortho}Upper: an orthogonalized tensor with respect to
    contraction over leg 1 and 2 and its adjoint. Lower: contracting the two
    symmetry tensors by summation over $m$ simplifies the contraction. $\{j_1\
    j_2\ j_3\}$ represents the triangle inequality.
  }
\end{figure}

A graphical depiction for this case is given in fig.~\ref{fig:leftortho}. In the
bottom of fig.~\ref{fig:leftortho} it is shown how the symmetry tensors simplify
the contraction. Indeed, with help of the symmetry tensors and the orthogonality
of $T$, one can show that
\begin{equation}
  \mathcal{C}(\mathbf{A} \mathbf{A}^+) = \sum_{\dofa{3}}|\dofa{3}\rangle
  \langle\dofa{3}|~.
\end{equation}

\begin{figure}[!ht]
  \centering
  \includegraphics[width=\columnwidth]{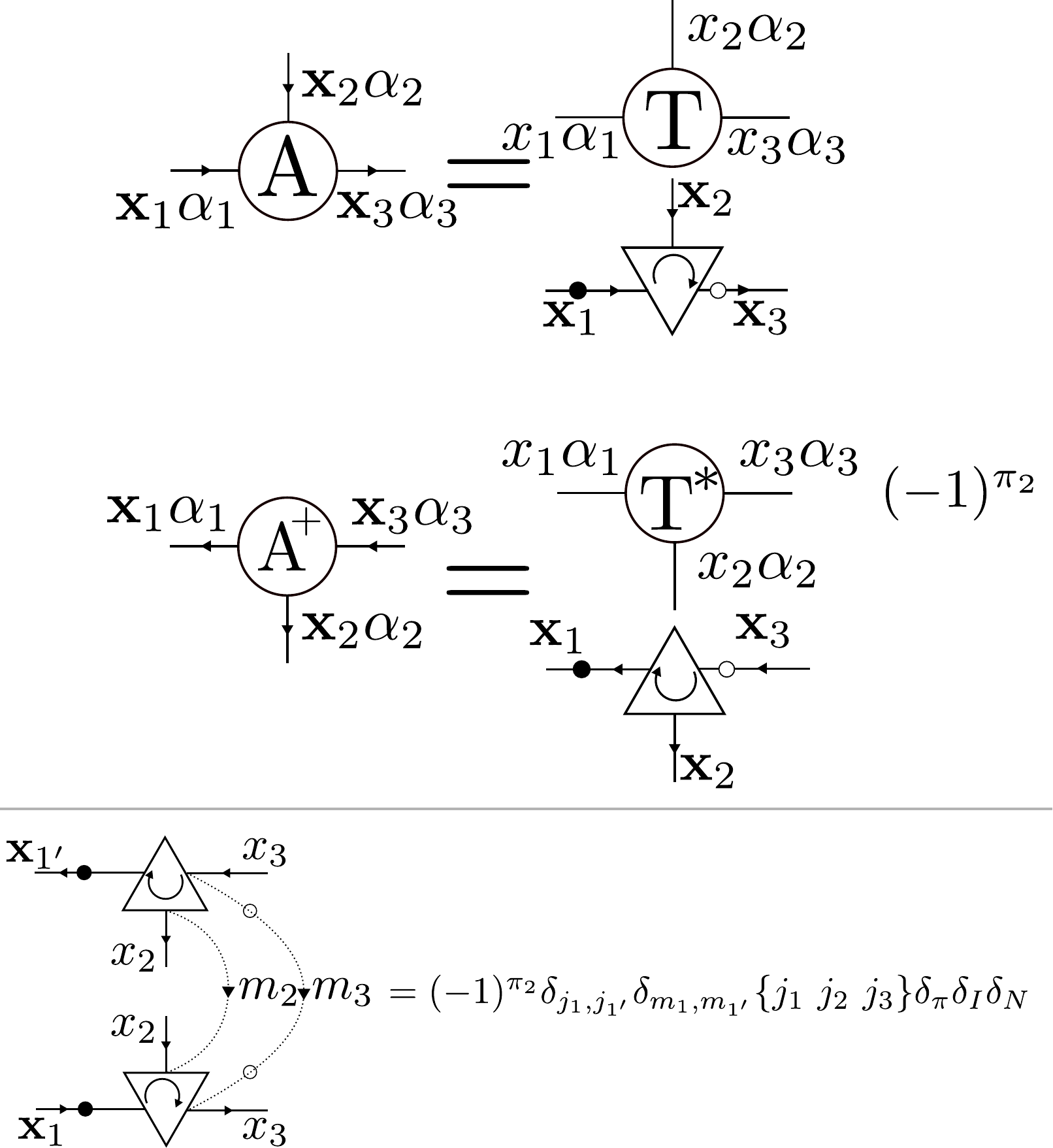}
  \caption{\label{fig:rightortho}Upper: an orthogonalized tensor with respect to
    contraction over leg 2 and 3 and its adjoint. Lower: contracting the two
    symmetry tensors by summation over $m$ simplifies the contraction. $\{j_1\
    j_2\ j_3\}$ represents the triangle inequality. The term $(-1)^{\pi_2}$ is
    introduced due to the fermionic nature of the tensors.
  }
\end{figure}

If the same reducible tensor has to be orthogonal with respect to a contraction 
over index 2 and 3, the reduced tensor should be again orthogonal with respect 
to that contraction. We use the gauge freedom to move a factor of $\sqrt{2j_1 + 
1}$ from a neighboring tensor to this tensor and a factor of $\sqrt{2j_3 + 1}$ 
from this tensor to a neighboring tensor (see the different position of the 
solid circle in fig.~\ref{fig:leftortho} and fig.~\ref{fig:rightortho}).

The adjoint of the reduced tensor is given again by its hermitian, but now, also
an extra phase needs to be introduced. This can be seen in
fig.~\ref{fig:rightortho}.  The phase is needed for imposing orthogonality due
to the fermionic nature of the tensors. The introduction of this phase does not
pose a problem as long the contraction over all adjoint tensors of the network
results in the expected bra wave function.  This can be ensured by correcting
all the introduced phases in the adjoint of the orthogonality center. We get
indeed for the orthogonalized tensor given in fig.~\ref{fig:rightortho} that
\begin{equation}
  \mathcal{C}(\mathbf{A} \mathbf{A}^+) = \sum_{\dofa{1}}|\dofa{1}\rangle
  \langle\dofa{1}|~.
\end{equation}

\subsection{Optimizing tensors}
\label{ap:optimize}
\begin{figure*}[!ht]
  \centering
  \includegraphics[width=\textwidth]{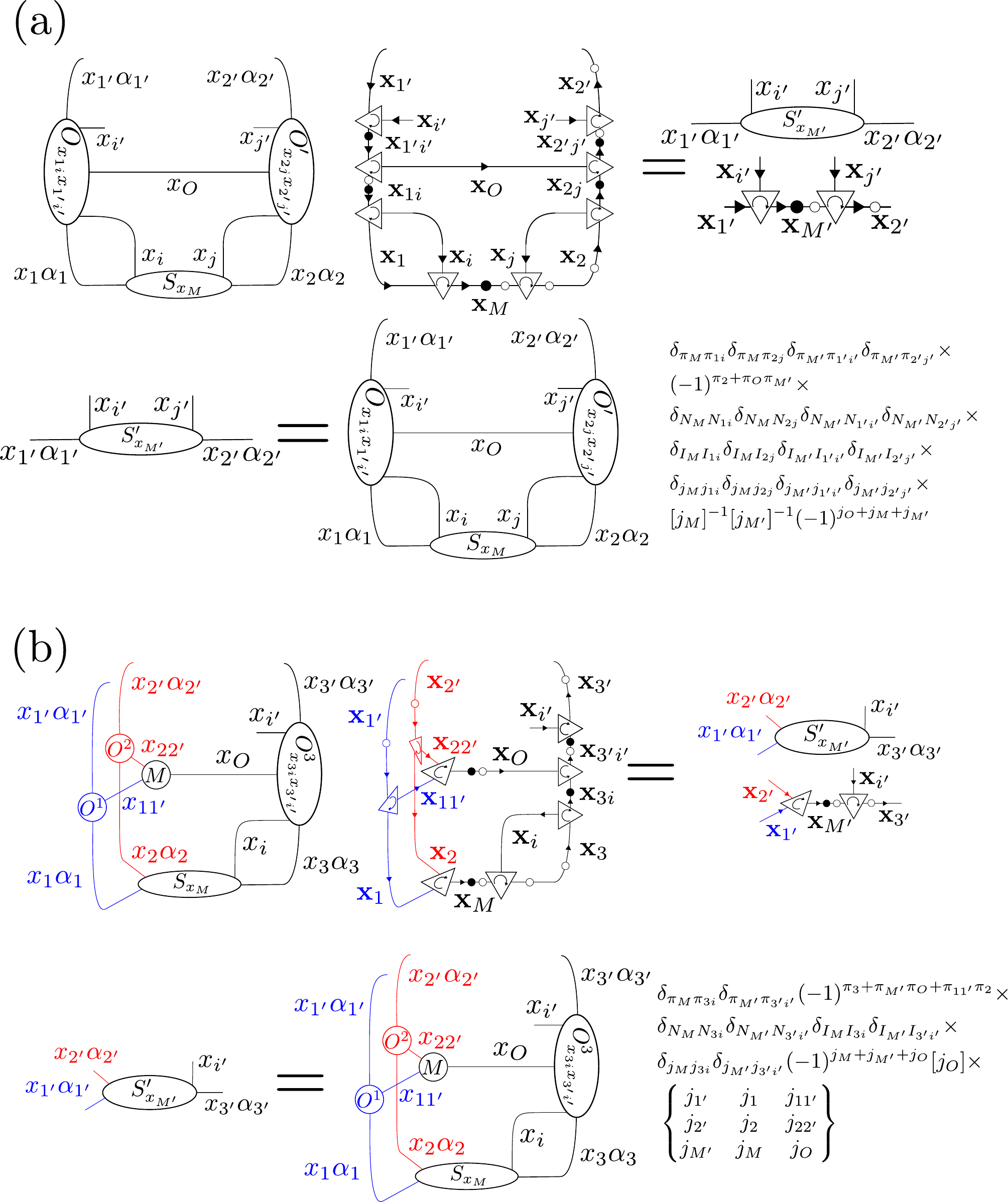}
  \caption{(a) The effect of the effective Hamiltonian on a two-site object
    consisting out of two physical sites. (b) The effective Hamiltonian on a
    two-site object consisting out of one physical and one branching site.
    Tensor $M$ is a tensor needed for the merging of the renormalized operators
    and its value can be found by deforming the network in fig.~\ref{fig:Vijkl}
    to the same form as the T3NS network.
  }
  \label{fig:heff}
\end{figure*}

The renormalized operators are used during the optimization of two contiguous
sites. Depending on the nature of the two sites, a different kind of
optimization is needed.  If both sites are physical, a DMRG-like optimization
step is used.  Two sets of renormalized operators are used to calculate the
effect of the effective Hamiltonian on the two-site object. This kind of
optimization is shown in fig.~\ref{fig:heff}(a). During this optimization step,
no Wigner symbols are needed. It also means that, in this formulation, one can
do an optimization of a DMRG-chain which only needs Wigner symbols when
appending site-operators to the renormalized operators.

If one of the two sites is a branching site, a T3NS optimization step is needed.
During this optimization, three sets of renormalized operators are used to
calculate the effect of the effective Hamiltonian on the two site object. An
example of this optimization step is shown in fig.~\ref{fig:heff}(b). Depending
on where the physical tensor is attached to the branching tensor, variants of
fig.~\ref{fig:heff}(b) are needed, giving rise to slightly different prefactors.

\subsection{Updating renormalized operators}
\label{ap:update}
After every optimization step of the algorithm the renormalized operators need
to be updated with the newly found optimized tensors. In
section~\ref{sec:optimization}, we already discussed appending a site-operator
to a renormalized operator (see fig.~\ref{fig:siteappend}). Here, we show how a
physical tensor or a branching tensor can be used to update the renormalized
operators.

\begin{figure}[!ht]
  \centering \includegraphics[width=\columnwidth]{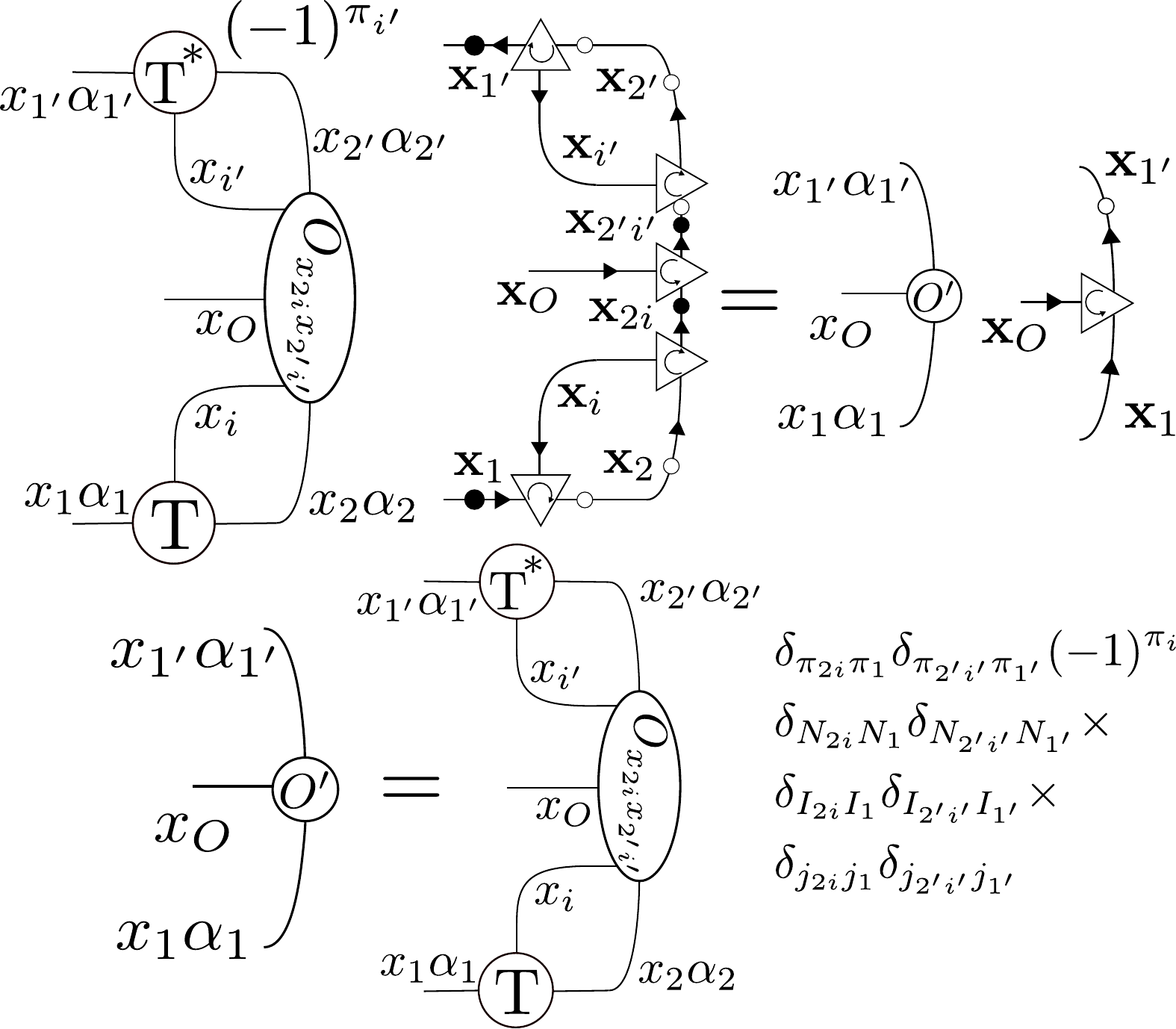}
  \caption{Update of a renormalized operator with a site-operator attached. This
    operator can be obtained as shown in fig.~\ref{fig:siteappend}. The update
    is done by using a orthogonalized physical tensor and its adjoint. In this
    example, the tensor is orthogonalized with respect to contraction over bond
    i and 2.
  }
  \label{fig:updateP}
\end{figure}

In fig.~\ref{fig:updateP}, a physical tensor is used to update a renormalized
operator with a site-operator appended to it. The physical tensor used in this
figure is orthogonal with respect to a contraction over leg 2 and 3 (see
fig.~\ref{fig:rightortho}).

\begin{figure}[!ht]
  \centering
  \includegraphics[width=\columnwidth]{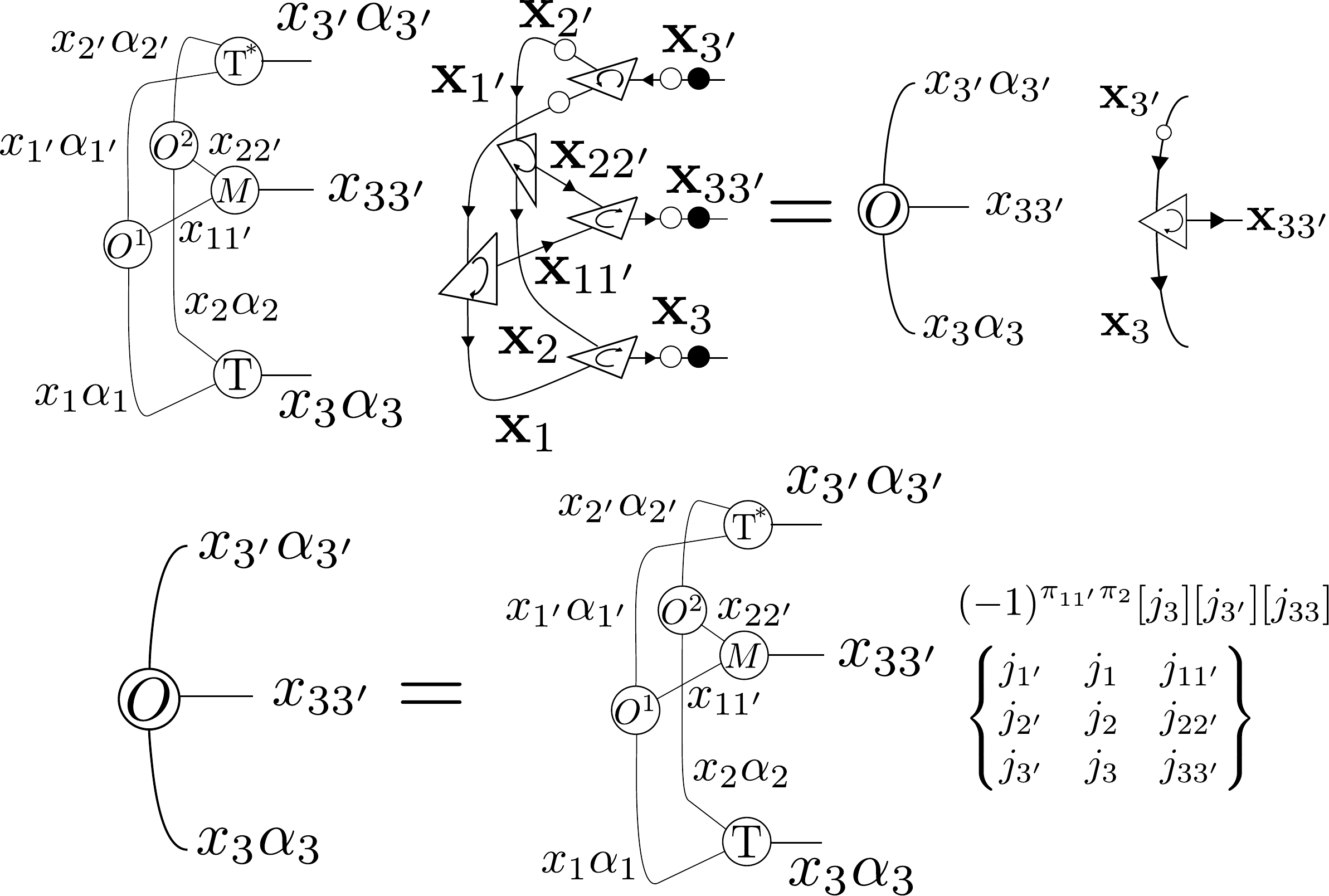}
  \caption{Recombination of two renormalized operators to a new one. The
    recombination is done by using an orthogonalized branching tensor and its
    adjoint. In this example, the tensor is orthogonalized with respect to
    contraction over bond 1 and 2. Tensor $M$ is a tensor needed for the merging
    of the two renormalized operators and its value can be found by deforming
    the network in fig.~\ref{fig:Vijkl} to the same form as the T3NS network.
  }
  \label{fig:updateB}
\end{figure}

In fig.~\ref{fig:updateB}, a branching tensor is used to recombine two
renormalized operators to a new one. The branching tensor in this example is
orthogonal with respect to a contraction over leg 1 and 2 (see
fig.~\ref{fig:leftortho}).

Similar graphical depictions can be made for both fig.~\ref{fig:updateP} and 
\ref{fig:updateB} when using tensors orthogonalized in different ways.
Both the appending of a site tensor and the recombination of two renormalized
operators through usage of a branching tensor need Wigner-9j symbols.  Updating of
a renormalized operator through usage of a physical tensor does not need any
Wigner symbols.

\bibliography{mybib}
\end{document}